\documentclass[]{aa520}
\usepackage{graphics,latexsym,epsfig,graphicx}     
\begin{document}
\title{Separating cosmic shear from intrinsic galaxy alignments:
correlation function tomography} 
\author{Lindsay J. King \& Peter Schneider} 
\institute{Institut f{\"u}r Astrophysik und Extraterrestrische Forschung, 
Universit{\"a}t Bonn, Auf dem
H{\"u}gel 71, D-53121 Bonn, Germany\\ } 
\date{} \authorrunning{Lindsay King \&
Peter Schneider} 
\titlerunning{Galaxy ellipticity correlation function
tomography} 
\abstract{During the past few years, secure
detections of cosmic shear have been
obtained, manifest in the correlation of the observed ellipticities of
galaxies. Constraints have already been placed on cosmological
parameters, such as the normalisation of the matter power spectrum 
$\sigma_{8}$. One possible systematic contaminant of the lensing 
correlation signal arises from intrinsic galaxy alignment, which is
still poorly constrained.  Unlike lensing, intrinsic correlations only
pertain to galaxies with small physical separations, the correlation 
length being a few Mpc. We present a new method that
harnesses this property, and isolates the lensing and intrinsic
components of the galaxy ellipticity correlation function using
measurements between different redshift slices. The observed signal is
approximated by a set of template functions, making no strong assumptions about
the amplitude or correlation length of any intrinsic alignment. We
also show that the near-degeneracy between the matter density
parameter $\Omega_{\rm m}$ and $\sigma_{8}$ can be lifted using 
correlation function tomography, even in the presence of an intrinsic 
alignment signal.
 }

\def\A{{\cal A}}
\def\eck#1{\left\lbrack #1 \right\rbrack}
\def\eckk#1{\bigl[ #1 \bigr]}
\def\rund#1{\left( #1 \right)}
\def\abs#1{\left\vert #1 \right\vert}
\def\wave#1{\left\lbrace #1 \right\rbrace}
\def\ave#1{\left\langle #1 \right\rangle}
\def\arcsecf {\hbox{$.\!\!^{\prime\prime}$}}
\def\arcminf {\hbox{$.\!\!^{\prime}$}}
\def\bet#1{\left\vert #1 \right\vert}
\def\vp{\varphi}
\def\vt{{\vartheta}}
\def\map{{$M_{\rm ap}$}}
\def\d{{\rm d}}
\def\mj{$\rm {m_{j}}$}
\def\mk{$\rm {m_{k}}$}
\def\col{$\rm {m_{j}}-\rm {m_{k}}$}\def\eps{{\epsilon}}
\def\vc{\vec} 
\def\s{{\rm d}}
\def\s{{\rm s}}
\def\t{{\rm t}}
\def\E{{\rm E}}
\def\L{{\cal L}}
\def\i{{\rm i}}
\def\seps{{\sigma_{\epsilon}}}
{\catcode`\@=11
\gdef\SchlangeUnter#1#2{\lower2pt\vbox{\baselineskip 0pt \lineskip0pt
  \ialign{$\m@th#1\hfil##\hfil$\crcr#2\crcr\sim\crcr}}}}
\def\gtrsim{\mathrel{\mathpalette\SchlangeUnter>}}
\def\lesssim{\mathrel{\mathpalette\SchlangeUnter<}}      

\maketitle
\section{Introduction}

The tidal gravitational field of mass inhomogeneities distorts the
images of distant galaxies, resulting in correlations in their
observed ellipticities. This cosmological weak lensing signal, or
cosmic shear, depends upon cosmological parameters and the matter
power spectrum (Blandford et al. 1991; Miralda-Escud\'e 1991; Kaiser
1992). In 2000, four teams announced the first detections of cosmic
shear (Bacon et al. 2000; Kaiser et al. 2000; van Waerbeke et
al. 2000; Wittman et al. 2000; Maoli et al. 2001), and more recently
measurements at arcminute scales have been made using the HST
(H\"ammerle et al. 2002; Refregier et al. 2002). Interesting
constraints have already been placed on the matter power spectrum
normalisation $\sigma_{8}$, and cosmic shear is also particularly
sensitive to the matter density parameter $\Omega_{\rm m}$, the power
spectrum shape parameter $\Gamma$ and the source redshift distribution
(e.g. van Waerbeke et al. 2002a; Hoekstra et al.\ 2002).  Future
multi-colour surveys will cover hundreds of square degrees, and have
the potential to place tight constraints on cosmological parameters
particularly when combined with results from the CMB, SNIa and galaxy
surveys (Mellier et al. 2002; van Waerbeke et al. 2002b). For example,
 van Waerbeke et al. (2002a) compared their constraints on
$\Omega_{\rm m}$ and $\sigma_{8}$ from lensing with those of Lahav et 
al. (2002) from the CMB, noting their near orthogonality.   
   
Various statistical measures of the cosmic shear have been suggested; here we 
focus on the two-point shear correlation function $\xi_{+}$ (hereafter
denoted by $\xi$), which is
convenient since it is insensitive to gaps in the data field, unlike
integrated measures such as the aperture mass statistic 
${\cal M}_{\rm ap}$ (e.g. Schneider et al. 1998). 

A possible systematic contaminant
of the lensing correlation function $\xi^{\rm L}$ is intrinsic 
alignment, which may arise during the galaxy formation process. This has been 
subject to numerical, analytic and 
observational studies [e.g. Croft \& Metzler 2000; Heavens et al. 2000 (HRH); 
Crittenden et al. 2001; Catelan et al. 2001; Mackey et al. 2002; Brown et
al. 2002; Jing 2002; Hui \& Zhang 2002], where amplitude estimates span a few 
orders of magnitude due to differences in the mechanism assumed to be 
responsible, and the type of galaxy considered. Nevertheless, these studies 
agree that the intrinsic correlation signal $\xi^{\rm I}$
can dominate the lensing signal for surveys with ${\ave z}\lesssim 0.5$.

Any correlation in ellipticities due to intrinsic alignment only
arises from physically close galaxy pairs, whereas $\xi^{\rm L}$ is
sensitive to the integrated effect of the density fluctuations out to
the redshift of the nearer galaxy. It has been shown that photometric
redshift information could be used to suppress $\xi^{\rm I}$, 
by downweighting or ignoring galaxy pairs at approximately the same
redshift (King \& Schneider 2002), or by downweighting nearby
pairs and subtracting a model of the
intrinsic alignment signal from the observed ellipticity correlation
function (Heymans \& Heavens 2002).

Motivated by the fact that intrinsic galaxy alignment is not yet well
understood, we present a new method to isolate the intrinsic and
lensing-induced components of the galaxy ellipticity correlation
function. This method assumes that photometric redshift information is
available, so that the correlation function can be measured between
different redshift slices. However, no specific model for intrinsic
ellipticity correlation (for instance its correlation length or
redshift evolution) needs to be adopted.  In the next section we
outline the method and in Sect.\,3 we present some results in the
context of a possible future survey. We discuss the results in
Sect.\,4.

\section{General method and specific assumptions}

In this section we outline a method to separate and extract the intrinsic and
lensing components of the galaxy ellipticity correlation function. The
general method is described in Sect.\,\ref{gen}, and in
Sect.\,\ref{spec} and Sect.\,\ref{basis} we state the assumptions
particular to this work.

\subsection{Method\label{gen}}
The ellipticity correlation function for galaxies with angular separation 
$\theta$ and at true redshifts $z_{i}$, $z_{j}$, is composed of a 
lensing and an intrinsic signal
\begin{equation}
\xi(\theta,z_{i},z_{j}) = \xi^{\rm L}(\theta,z_{i},z_{j}) + \xi^{\rm I}(\theta,z_{i},z_{j})\;.
\end{equation}
As noted in Sect.\,1, the origin and behaviour of $\xi^{\rm I}$ is
not yet well understood. $\xi^{\rm L}(\theta,z_{i},z_{j})$ is related to the 3-dimensional 
matter power spectrum $P_\delta$ through
\begin{eqnarray}
&&\xi^{\rm L}(\theta,z_i,z_j)=
{9 H_0^4 \Omega_{\rm m}^2\over 4 c^4}
\int_0^{{\rm min}[w_{i},w_{j}]} {\d w\over a^2(w)}~~~~~~\nonumber \\
&&\times R(w,w_{i})\,R(w,w_{j}) 
 \int{\d \ell\,\ell\over (2\pi)}\,P_\delta\rund{{ \ell\over f(w)},w}\,{\rm
J}_0(\ell\theta) \,,
\label{powspec}
\end{eqnarray}
where $H_0$ and $\Omega_{\rm m}$ are the values of the Hubble
parameter and matter density parameter at the present epoch, and $a(w)$ is
the scale factor at comoving distance $w$, normalised such that
$a(0)=1$ today. J$_{0}$ is the $0$-th
order Bessel function of the first kind, and $\vec\ell$ is the angular
wave-vector. We denote the comoving distance at $z_{i}$ by 
$w_{i}$, and the function 
$f(w)$ is the comoving angular diameter distance, which depends on the
spatial curvature $K$:
\begin{equation}
  f(w) = \left\{
  \begin{array}{ll}
    K^{-1/2}\sin(K^{1/2}w) & (K>0)\\
    w & (K=0)\\
    (-K)^{-1/2}\sinh[(-K)^{1/2}w] & (K<0) \\
  \end{array}\right\}\;.
\end{equation}
The function $R(w,w')=f(w'-w)/f(w')$ is the ratio
of the angular diameter distance of a source at comoving distance $w'$
seen from a distance $w$, to that seen from $w=0$. 

Next, we account for the availability of photometric redshift
estimates rather than spectroscopic ones. The galaxy ellipticity correlation function 
becomes  
\begin{equation}
{\bar \xi}(\theta,{\bar z_{i}},{\bar z_{j}})=\int{\rm d}z_{i}\int{\rm
d}z_{j}\,p(z_{i},z_{j}|{\bar z_{i}},{\bar z_{j}},\theta)\,\xi(\theta,z_{i},z_{j})\;,
\label{phot}
\end{equation}
where $p(z_{i},z_{j}|{\bar z_{i}},{\bar z_{j}},\theta)$ is 
the probability to have true redshifts $z_i$ and $z_j$ given
photometric estimates ${\bar z_{i}}$ and ${\bar z_{j}}$. This is given by
\begin{eqnarray}
&&p(z_{i},z_{j}|{\bar z_{i}},{\bar z_{j}},\theta)=\\\nonumber
&&\frac{p(z_{i}|{\bar z_{i}})\,p(z_{j}|{\bar z_{j}})\,[1+\xi_{\rm gg}(r)]}{\int{\rm d}z_{i}\int{\rm d}z_{j}\,p(z_{i}|{\bar z_{i}})\,p(z_{j}|{\bar z_{j}})\,[1+\xi_{\rm gg}(r)]}\;,
\end{eqnarray}
where $\xi_{\rm gg}$ is the galaxy spatial correlation function, which
may also include redshift dependent evolution, and $r$ is the comoving separation of the galaxies in a pair. 

We now assume that ${\bar \xi}(\theta,{\bar z_{i}},{\bar z_{j}})$ is 
available on a 3-dimensional grid of $N_{K}$ angular separation bins of 
width $\Delta\theta$ centred on $\theta_{K}$ (index $K$), and 
$N_{Z}$ photometric redshift bins of width $\Delta z$ centred on each of
${\bar z_{i}}$ (index $I$) and ${\bar z_{j}}$ (index $J$). This could
either correspond to an observed signal ${\bar \xi}^{\rm obs}_{IJK}$, or to a
theoretical prediction $\ave{{\bar \xi}^{\rm mod}}_{IJK}$ 
which we want to compare with the observed signal. 

We assume that both the lensing and intrinsic correlations can be written in terms of
sets of template functions $A_{n}$ and $B_{n}$
\begin{eqnarray}
&&\bar {\xi^{\rm L}}(\theta,{\bar z_{i}},{\bar z_{j}})=\sum_{n=1}^{{\rm N_{L}}}a_{n}A_{n}(\theta,{\bar z_{i}},{\bar z_{j}})\;,\\\nonumber
&&\bar {\xi^{\rm I}}(\theta,{\bar z_{i}},{\bar z_{j}})=\sum_{n=1}^{{\rm N_{I}}}b_{n}B_{n}(\theta,{\bar z_{i}},{\bar z_{j}})\;,
\end{eqnarray}
where $a_{n}$ and $b_{n}$ are the amplitudes of the $n$-th
lensing and $n$-th intrinsic template functions.
The template functions are fairly arbitrary, and extra functions can be added 
as required, to span the range of plausible models. We describe our
choice of models in Sect.\,\ref{basis} below.

A suitably chosen single index $m$ identifies correlations between
bins with redshift indices $I$,$J$ and angular separation index $K$.
In total, there are $N_{M}=N_{Z}\,(N_{Z}+1)\,N_{K}/2$ such independent
measurements. The total of $N=N_{L}+N_{I}$ gridded template models for
the correlation functions ($A_{1}...A_{N_{L}}, B_{1}...B_{N_{I}}$)
can be written as an $N_{M}\times N$ so-called design matrix ${\cal
M}$, and their amplitudes ($a_{1}...a_{N_{L}}, b_{1}...b_{N_{I}}$) as
an $N$-dimensional column vector ${\cal G}$ so that
\begin{equation}
\ave{{\bar\xi}^{\rm mod}}_{m}={\cal M}_{mn}{\cal G}_{n}\;.
\end{equation}
Our aim is to recover $\bar{\xi}^{\rm obs}_{m}$ in terms of the template
functions. Using the method of least squares, the best-fit 
estimates of the ${\cal G}_{n}$ are those values ${\hat {\cal G}}_{n}$ 
which minimise
\begin{equation}
S=(\vec{\bar{\xi}}^{\rm obs}-{\cal M}{\cal G})\,{\cal C}^{-1}\,(\vec{\bar{\xi}}^{\rm
obs}-{\cal M}{\cal G})\;,
\end{equation} 
where ${\cal C}$ is the covariance matrix. Since $\ave{\bar{\xi}^{\rm mod}}_{m}$
is a linear combination of template functions, the linear least squares estimators are 
\begin{equation}
{\hat {\cal G}}=\left({\cal M}^{T}{\cal C}^{-1}{\cal
M}\right)^{-1}{\cal M}^{T}{\cal C}^{-1}\vec{\bar{\xi}}^{\rm obs}\;
\label{rec}
\end{equation}
After obtaining ${\cal C}$, combined with an observed correlation function 
$\vec{\bar{\xi}}^{\rm obs}$ and a design matrix ${\cal M}$, we have the
necessary machinery to obtain ${\hat {\cal G}}$ i.e. the projection of $\vec{\bar{\xi}}^{\rm obs}$
into the template functions. This means that separate fits for the 
lensing and intrinsic contributions can be obtained.     

\subsection{Covariance matrix\label{spec}}
To evaluate (\ref{rec}), we need the covariance matrix ${\cal
C}_{mm'}$. The covariance matrix could be calculated using the method
described in Schneider et al.\ (2002), where ${\cal C}$ is expressed
as integrals over (products of) correlation functions. However, since
the method presented here would require the calculation of very many
elements of the covariance matrix, owing to the redshift slicing, we
decided to use, as a first step, a simplified model for ${\cal C}$.
This consists of neglecting the cosmic variance contribution to ${\cal
C}$, and thus consideration of the (diagonal) elements of ${\cal C}$ coming
from the intrinsic ellipticity dispersion of the source galaxies.
This is the dominant contribution to the covariance at small angular
scales; at larger angular scales, the cosmic variance terms start to
dominate, with the transition angular scale depending on the survey
geometry (Kaiser 1998; Schneider et al.\ 2002). Here, we consider
${\cal N_{F}}$ independent fields, and take ${\cal
N_{F}}=300$. Therefore, we expect the cosmic variance not to be very
much larger than the intrinsic ellipticity noise on the angular scales
considered. 

\begin{figure}[width=88mm]
\epsfig{file=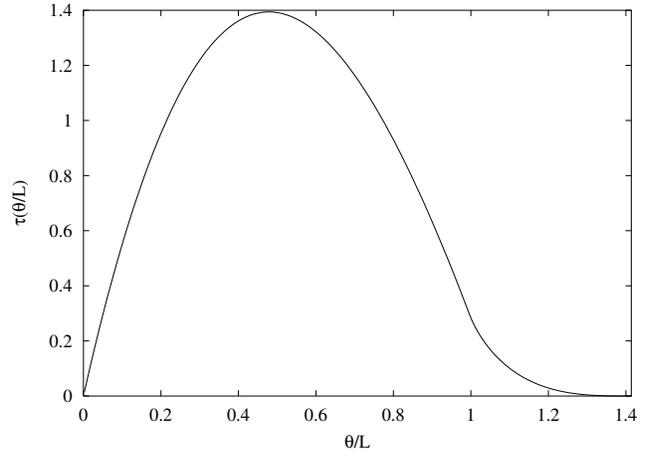, width=88mm}
\caption{The function $\tau(\theta/L)$ (as defined in the text) which 
accounts for the finite extent of the data field when calculating the
number of pairs in a given bin.}        
\label{tau}
\end{figure}

With this approximation, the elements of the covariance matrix are 
\begin{equation}
{\cal C}_{m\,m'}=
\frac
{2\left(\seps^{2}/2\right)^{2}}
{{\cal N_{\rm p}}(m)}\left[
\delta_{m,m'}
\left(1\,+\,\delta_{I,I'}\delta_{J,J'}\delta_{I,J}
\right)
\right]\;,
\end{equation}
where there is an extra contribution from auto-variance terms. The
elements of the inverse covariance matrix in this case are simply 
$\left({\cal C}^{-1}\right)_{m\,m'}=\delta_{m\,m'}/{\cal C}_{m\,m}$.
The galaxy ellipticity dispersion is denoted by $\seps$. In bin $m$,
the number of pairs is given by
\begin{equation}
{\cal N_{\rm p}}(m)={\cal N_{\rm F}}\,\left[n_{0}p(z_{I})\Delta
z\right]\,\left[n_{0}p(z_{J})\Delta z\right]L^{4}\frac{\Delta\theta}{L}\tau\left(\frac{\theta}{L}\right)\;,
\end{equation} 
where $n_{0}$ is the galaxy 
number density, $p(z_{I})$ is the redshift probability density for redshift 
bin $I$, and $L$ is the extent of the field, assumed to be square. 
$\tau(\theta/L)$ is a function that 
takes into account the fact that fields have finite extent, and it must be evaluated numerically.
Fig.\,\ref{tau} shows the function $\tau(\theta/L)$; note the
limiting case where pairs separated by more than $\sqrt 2{L}$ do not occur.

\subsection{Basis models\label{basis}}
In order to illustrate the general method described above, we will 
choose a simple, restricted set of template functions which share the
approximate functional behaviour expected from the real correlation
functions, both intrinsic and lensing. For the latter, we simply take
a small number of CDM cosmologies and consider their correlation
functions as a template set. For the former, simple exponentials with
redshift dependence are chosen.

The lensing template functions $A_{n}$ used here are the gridded
${\bar \xi^{\rm L}}(\theta,{\bar z_{i}},{\bar z_{j}})$ for 3 models of
the underlying cosmology: $\Lambda$CDM ($\Omega_{\rm m}=0.3$, 
$\Omega_{\Lambda}=0.7$),
OCDM ($\Omega_{\rm m}=0.3$, $\Omega_{\Lambda}=0$) and $\sigma$CDM
($\Omega_{\rm m}=1$, $\Omega_{\Lambda}=0$); for simplicity, 
$\sigma_{8}=0.9$ and $\Gamma=0.21$ for each model. We use the Bardeen
et al. (1986) transfer function to describe the evolution of the
3-dimensional power spectrum, along with the prescription of Peacock \& Dodds (1996) 
for evolution in the non-linear regime. The required lensing correlation 
functions are calculated using the relationship between the power spectrum
and $\xi^{\rm L}$ given in (\ref{powspec}), and then integrated over the
photometric redshift uncertainties as in (\ref{phot}).
Here it is assumed that $p(z|{\bar z})$ is a Gaussian with dispersion
$\sigma_{\rm phot}$, centred on ${\bar z}$. 

Nine template models $B_{n}$ for the intrinsic alignments are
considered. First, the
true spatial intrinsic correlation function is parameterised in terms
of a correlation length $R_{\rm corr}$ and an exponent $\alpha$:   
\begin{equation}
\eta(r,z)=(1+z_{\rm av})^{\alpha}\left[{\rm exp}\,\left(-r/R_{\rm corr}\right)\right]\;,
\end{equation}
where $z_{\rm av}$ is the mean redshift of galaxies in a
pair and $r$ is their comoving separation. We use the approximation 
$r^{2}=(w_{i}-w_{j})^{2}\,+\,\theta^{2}f^{2}\left[(w_{i}+w_{j})/2\right]$.
$R_{\rm corr}$ was taken to be [1, 3, 10]$\,h^{-1}\,{\rm Mpc}$
and $\alpha$ to be [$-$1, 0, 1]. The availability of photometric
redshift estimates is then accounted for by integrating $\eta(r,z)$
as in (\ref{phot}), and finally we obtain each of the model
correlation functions on a grid. Note that the intrinsic models are calculated
using the relationships for the distances $f(w)$ pertaining to the $\Lambda$CDM cosmology 
described above, and that we use 
$\xi_{\rm gg}(r)=(r/5\,h^{-1}\,{\rm Mpc})^{-1.8}$. 
\begin{figure*}
\includegraphics[width=12cm]{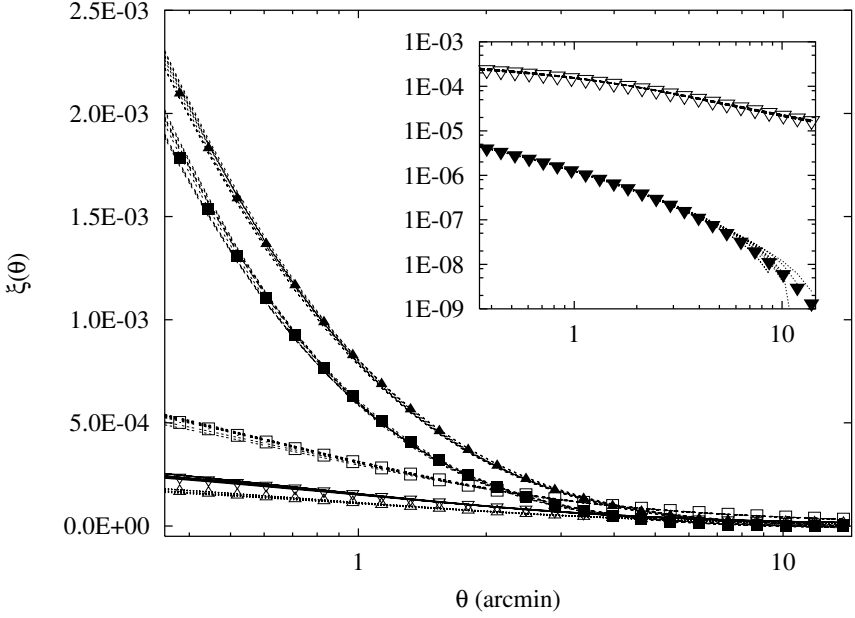}
\caption{The lensing and intrinsic cross-correlation functions
measured between three pairs of photometric redshift bins for the $\Lambda$CDM cosmology and HRH 
intrinsic alignment model. Open (filled) squares correspond to the
true lensing (intrinsic) signal for bins centred on ${\bar z_{i}}=1.07$
and ${\bar z_{i}}=1.13$. Open (filled) triangles show the true lensing
(intrinsic) signal for bins centred on ${\bar z_{i}}=0.62$
and ${\bar z_{i}}=0.68$. Open (filled) inverted triangles show the true
lensing (intrinsic) signal between bins centred on ${\bar z_{i}}=0.62$
and ${\bar z_{i}}=1.13$; since the intrinsic signal is so low, this is
plotted on the inlay panel along with the lensing signal. 
The sets of lines associated with each of the true model
symbols are the recovered best-fit lensing (intrinsic) cross-correlation 
functions to noisy realisations of $\bar{\xi}^{\rm obs}_{m}$ for
the pairs of photometric redshift bins, using the procedure described in the
text.
}        
\label{lcdm}
\end{figure*}

\begin{figure*}
\includegraphics[width=12cm]{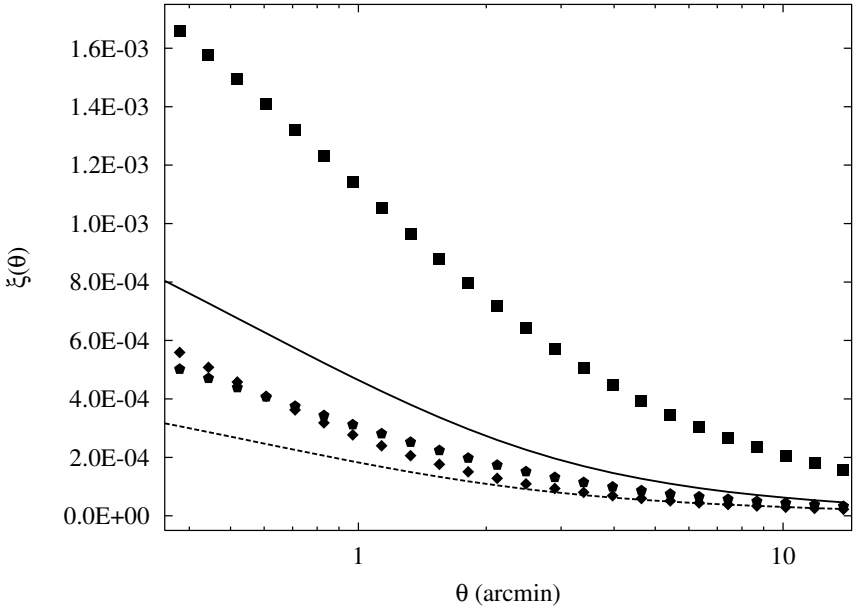}
\caption{The lensing correlation functions for the three cosmologies
used for the construction of template functions, and for the other cosmologies considered in Sect.\,3.1, plotted for the bins centred on ${\bar z_{i}}=1.07$
and ${\bar z_{i}}=1.13$. Filled squares correspond to 
$\sigma{\rm CDM}$, the filled pentagons indicate $\Lambda{\rm CDM}$ and filled 
diamonds correspond to OCDM. The dashed and solid lines 
indicate the flat cosmologies with (i) $\sigma_{8}=0.71$,
$\Omega_{\rm m}=0.33$ and $\Gamma=0.215$, and (ii) $\sigma_{8}=0.8$,
$\Omega_{\rm m}=0.5$ and $\Gamma=0.3$, respectively.
}
\label{cosmo}
\end{figure*}

\begin{figure*}
\includegraphics[width=12cm]{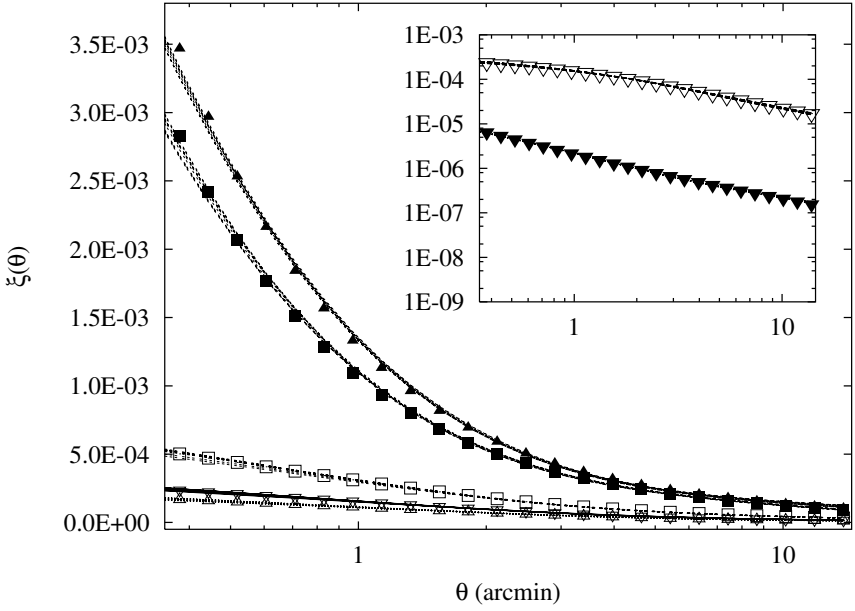}
\caption{The lensing and intrinsic correlation functions between
different redshift bins for the $\Lambda$CDM cosmology and Jing
(2002) intrinsic alignment model. The symbols and lines have the same
meaning as in Fig.\,\ref{lcdm}.}
\label{jing}
\end{figure*}
 
\begin{figure*}
\includegraphics[width=12cm]{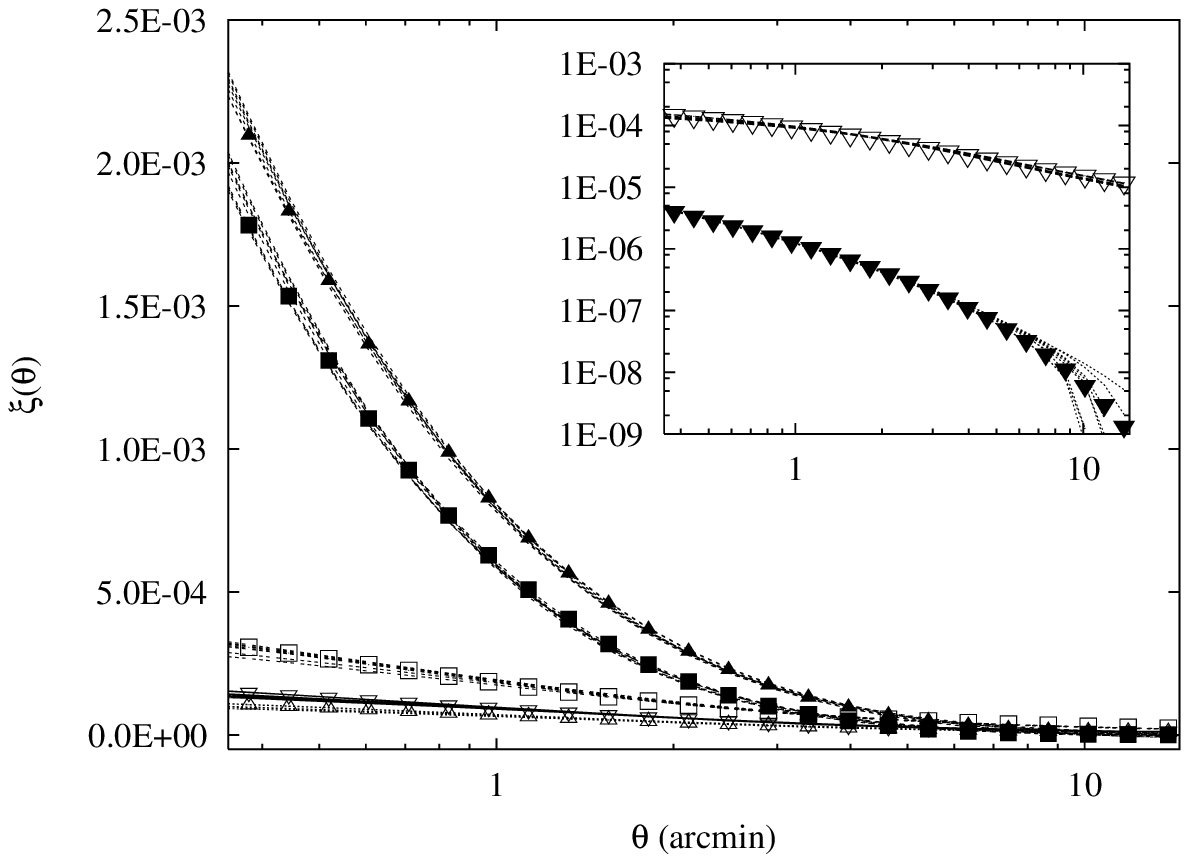}
\caption{The lensing and intrinsic correlation functions between
different redshift bins for a flat cosmology with $\Omega_{\rm m}=0.33$, $\sigma_{8}=0.71$ and $\Gamma=0.215$ and an 
intrinsic alignment model from HRH. The symbols and lines have the
same meaning as in Fig.\,\ref{lcdm}.}
\label{fake}
\end{figure*}

\begin{figure*}
\includegraphics[width=12cm]{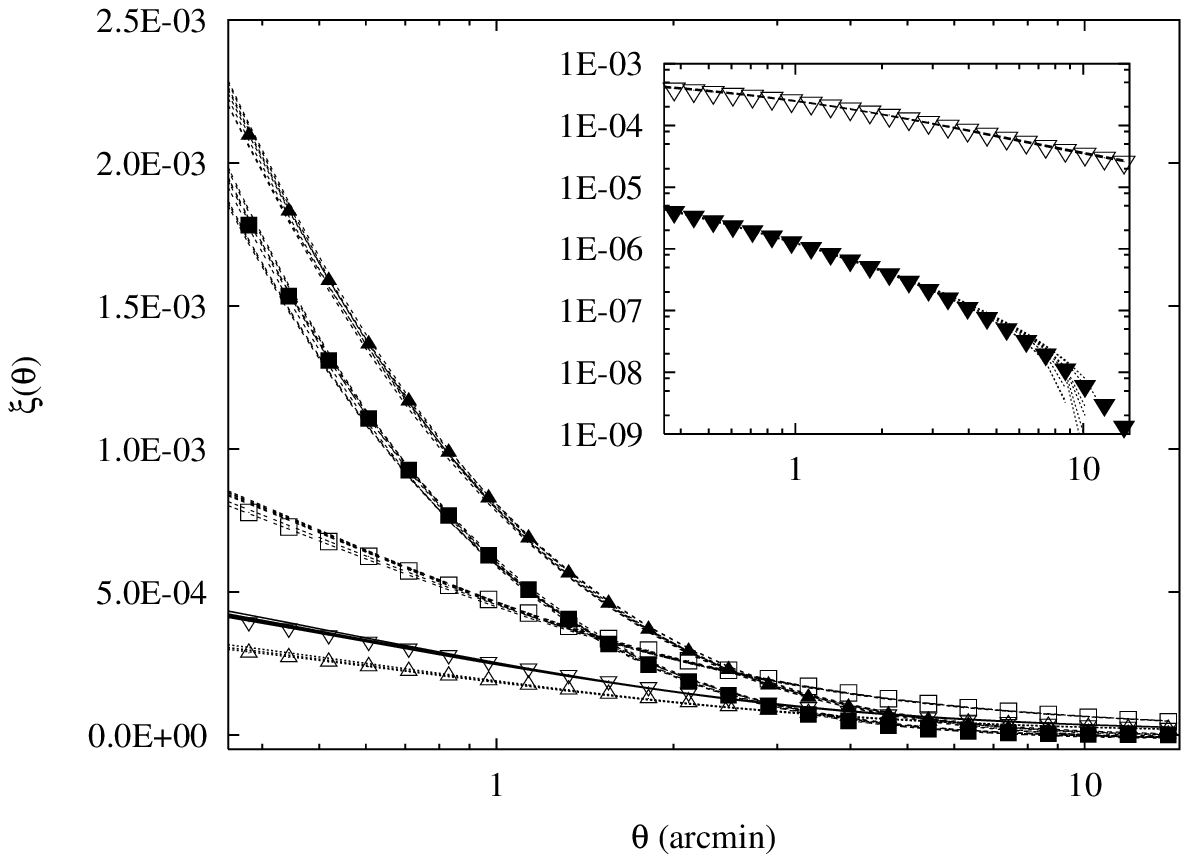}
\caption{The lensing and intrinsic correlation functions between
different redshift bins for a flat cosmology with $\Omega_{\rm m}=0.5$, $\sigma_{8}=0.8$ and $\Gamma=0.3$ and an 
intrinsic alignment model from HRH. The symbols and lines have the
same meaning as in Fig.\,\ref{lcdm}.}
\label{fake2}
\end{figure*}

\section{Results}
The results are presented in the context of a possible future
multi-colour cosmic shear survey, with a field size 
$L=14\arcmin$, and ${\cal N_{F}}=300$ independent pointings
(i.e. the largest scale on which the ellipticity correlation function
is available is $\sqrt 2\times 14\arcmin$). A galaxy
number density of 30\,arcmin$^{-2}$ and ellipticity dispersion of
$\seps=0.3$ are used throughout. The value of $\sigma_{\rm phot}=0.1$ is chosen, typical of that obtained with current SED fitting procedures such as
hyper-Z using a wide range of optical and near-infrared filters
(Bolzonella et al.\, 2000). There are $N_{Z}=65$ redshift slices between
${\bar z}=0.2$ and 2.12, and $N_{K}=25$ angular separation bins
between $0\arcminf 3$ and $15\arcmin$.

The galaxy redshift distribution follows
the parameterisation suggested by Smail et al. (1995),
i.e. $p({\bar z})=\beta/[z_{0}\,\Gamma_{\beta}(3/\beta)]({\bar z}/z_{0})^{2}\,{\rm
exp}\left(-({\bar z}/z_{0})^{\beta}\right)$, where $\Gamma_{\beta}$
denotes the gamma
function. We take $\beta=3/2$ and $z_{0}=2/3$ yielding ${\ave {\bar z}}\approx 1$.

Two applications of the technique are presented. We start by asking how
well the intrinsic and lensing contributions to the galaxy ellipticity correlation
function can be separated, using the information contained in 
correlation functions between different redshift slices. Secondly, 
it has been shown that the degeneracy between $\Omega_{\rm m}$ and
$\sigma_{8}$ can be partly lifted when redshift estimates for source
galaxies are available (e.g. van Waerbeke et al. 2002a). We illustrate 
the use of correlation function tomography in this respect. 

\subsection{Isolating the intrinsic correlation signal}

Now we consider input ``observed" correlation functions
$\bar{\xi}^{\rm obs}_{m}$ comprising a lensing and an intrinsic
contribution, to see how well the individual signals are recovered in 
terms of template functions using (\ref{rec}). First 
it was checked that in the absence of noise, 
${\hat {\cal G}}_{n}\equiv {\cal G}_{n}$, when $\vec{\bar{\xi}}^{\rm
obs}$ is composed of a lensing and an
intrinsic model contained in the set of templates. 

The intrinsic alignment model for spirals from HRH, 
$\eta(r,z)=0.012\,{\rm exp}(-r/1.5\,h^{-1}{\rm Mpc})$ was then used to 
obtain $\bar{\vec{\xi}^{\rm I}}$, and $\bar{\vec{\xi}^{\rm L}}$ was calculated for a 
$\Lambda$CDM cosmology. Random gaussian distributed errors with dispersion 
$\sigma={\cal C}_{mm}^{0.5}$ were added to these correlation functions 
giving noise realisations, and best-fit parameters ${\cal G}_{n}$ were
recovered for each of these. Fig.\,\ref{lcdm} shows the (noise-free) input and 
recovered intrinsic and lensing correlation functions between three 
combinations of redshift slices: one close pair at ${\bar z}\sim 1$, one close
pair at ${\bar z}\sim 0.6$, and the correlations between slices at
${\bar z}\sim 1$
and ${\bar z}\sim 0.6$. The lensing correlation function for each
of the three cosmologies used in the construction of template
functions is shown in Fig.\,\ref{cosmo}, plotted for
the slices at ${\bar z}\sim 1.0$. The intrinsic correlation signal surpasses 
the lensing signal out to several arcminutes for both the low-redshift
and the high-redshift bins. 
Considering bins with a large separation in redshift
reduces the intrinsic signal to a negligible level, as expected.
Even with our limited set of template functions, the reduced $\chi^{2}$ 
values of the recovered fits to the noise realisations are 
$\approx 1$. Also note that the intrinsic signal can be well represented
in terms of the template functions, although it is not contained in the
template set. 
 
Since the current template set for the intrinsic alignment signal
contains exponentially-decaying models with different scale-lengths,
we now consider how well the method fares if the true signal is a 
power-law instead. The intrinsic alignment model is taken from Jing
(2002); we use $\eta(r,z)=0.288/[r^{0.4}(7.5^{1.7}+r^{1.7})]$, with
$r$ measured in units of $h^{-1}\,{\rm Mpc}$. Again, 
the true cosmology is $\Lambda$CDM. Noise was added and best-fit 
parameters recovered in the same manner as described above. 
Fig.\,\ref{jing} shows the (noise-free) input and  
recovered intrinsic and lensing correlation functions between the same
three combinations of redshift slices as for Fig.\,\ref{lcdm}.
Even though the functional form of the true intrinsic signal is quite
different from the template models, the best-fit intrinsic models are still rather
close to the noise-free model and more importantly, the lensing signal is again
well recovered. To assess the difference between using the Jing model rather
than the HRH model for intrinsic alignments, the reduced $\chi^{2}$
values of the recovered fits were determined for 1000 noise
realisations of each, keeping the same $\Lambda$CDM cosmology. 
The reduced $\chi^{2}$ value is lower for the HRH 
realisation in nearly all cases, since this is an
exponentially-decaying model for which the templates are better
adapted. The $\chi^{2}$ values for the HRH realisations closely follow
the theoretically expected distribution. Although the intrinsic models can be distinguished
statistically, the difference in the mean $\chi^{2}$ values of the two
sets of 1000 realisations is only $\approx 15\%$ of their dispersion. In practice, several 
families of functional forms could be taken for the template set.

The next example for $\vec{\bar{\xi}}^{\rm obs}$ again uses the HRH
model for intrinsic alignments,
but this time a different flat cosmology with $\sigma_{8}=0.71$,
$\Omega_{\rm m}=0.33$, $\Gamma=0.215$ was used for the lensing
correlations; the lensing signal for this cosmology was not part of
 the template set. Fig.\,\ref{cosmo} shows the lensing
correlation function for this cosmology, plotted for the the slices at 
${\bar z}\approx 1.0$. Noise was added using the same random seeds as above, 
and best-fit parameters recovered as
before. Fig.\,\ref{fake} shows results for the same combination of
redshift slices as for Fig.\,\ref{lcdm}. The power spectrum
corresponding to this cosmology has a lower normalisation -- the
$\sim 20\%$ reduction in $\sigma_{8}$ is not offset by the $10\%$
increase in $\Omega_{\rm m}$; this is evident in the lower amplitude of
the lensing correlation functions. 
Again, the recovered parameters are consistent
with the true ``observed" correlation functions and the reduced $\chi^{2}$
values $\approx 1$.

Whereas the foregoing cosmological model was quite different in 
{\it amplitude} of the power spectrum we now consider a model which
also differs in the
{\it shape} of the power spectrum, to test the robustness of our
method. Keeping the HRH model for intrinsic alignments, a flat cosmology 
with $\sigma_{8}=0.8$,
$\Omega_{\rm m}=0.5$, $\Gamma=0.3$ was next used for the lensing
correlations. Fig.\,\ref{cosmo} shows the lensing correlation function
for this cosmology, plotted for the the slices at ${\bar z}\sim 1.0$.
Noise was added and best-fit parameters recovered
as described above. Fig.\,\ref{fake2} shows results for the same combination of
redshift slices as for Fig.\,\ref{lcdm}. The lensing signal is again well
represented in terms of the basis functions, even though it is rather
different to any of the cosmology templates. Hence, despite the fact
that our set of template functions is quite
restrictive, we have demonstrated that it provides enough flexibility
to provide accurate fits to the correlation functions of quite
different comological models.

\subsection{Breaking the $\Omega_{\rm m}$-$\sigma_{8}$ degeneracy}

\begin{figure}
\epsfig{file=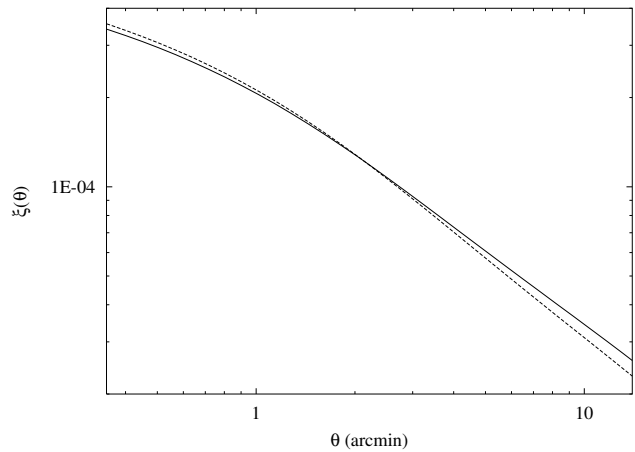, width=88mm}
\caption{This figure shows the lensing correlation functions for our
fiducial $\Lambda$CDM model: $\Omega_{\rm m}=0.3$, $\sigma_{8}=0.9$
(dashed line), and for a degenerate model with $\Omega_{\rm m}=0.4$,
$\sigma_{8}=0.78$ (solid line). Both functions were predicted using 
the matter power spectrum described in Sect.\,\ref{basis}, and for a
redshift distribution with ${\ave z}=1$.}
\label{degen}
\end{figure}

The near degeneracy, in the absence of redshift information, between
$\Omega_{\rm m}$ and $\sigma_{8}$ for two example flat cosmologies is
illustrated in Fig.\,\ref{degen}, taken from King \& Schneider (2002;
where details for its calculation can be found).  One cosmology is the
fiducial $\Lambda$CDM model: $\Omega_{\rm m}=0.3$, $\sigma_{8}=0.9$,
and the other is an almost degenerate model with $\Omega_{\rm m}=0.4$
and $\sigma_{8}=0.78$.  To obtain these correlation functions, we
assume the same prescription for the power spectrum outlined in
Sect.\,\ref{basis} above. The source population has a redshift
probability distribution with ${\ave z}=1$, but with no individual
photometric redshift estimates assumed to be available.

\begin{figure}
\epsfig{file=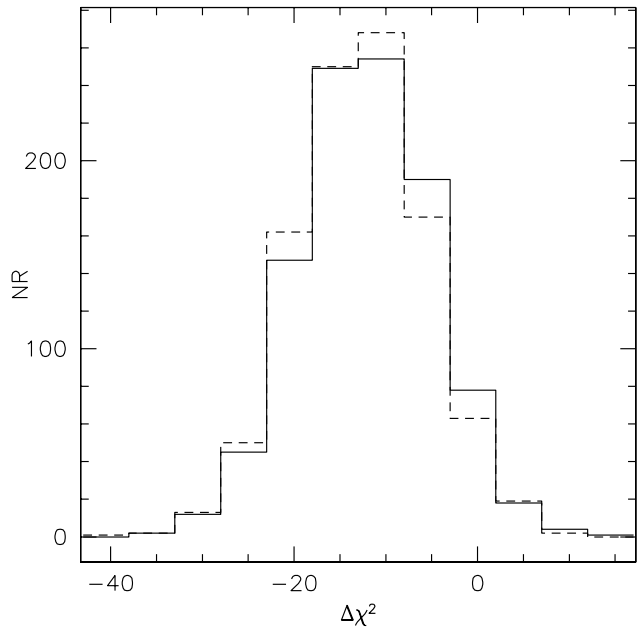, width=88mm}
\caption{These histograms show the frequency (NR) of values of 
$\Delta\chi^{2}$ for two
sets of 1000 noise realisations. The first (second) set of 
realisations indicated by the solid (dashed) histogram has the fiducial
(degenerate) $\Lambda$CDM cosmology in the template set and
$\Delta\chi^{2}$ is calculated between the best-fit $\Lambda$CDM
(degenerate) as opposed to degenerate ($\Lambda$CDM) recovered
model.}
\label{chisq}
\end{figure}

In order to see how well the fiducial and degenerate 
models could be distinguished using correlation function tomography, 
we took the gridded correlation functions for each model and added
each of these to the gridded HRH intrinsic correlation function, giving
$\ave{{\bar\xi}^{\rm mod}}_{m}$ for each. For simplicity, we refer to
these combinations as ${\vec{\bar\xi}}_{\Lambda\rm CDM}$ and
${\vec{\bar\xi}}_{\rm degen}$.  A first set of simulations involved
using a set of ten template functions containing the nine models for
intrinsic alignments, along with the $\Lambda$CDM lensing model for
the lensing template. In turn, 1000 noise realisations of
${\vec{\bar\xi}}_{\Lambda\rm CDM}$ and ${\vec{\bar\xi}}_{\rm degen}$
were generated using the same random seeds in both cases, and the best-fit
amplitudes ${\hat{\cal G}}_{n}$ for the template functions
recovered. The process was repeated, this time using the gridded
degenerate model as the lensing template function, in place of the
fiducial $\Lambda$CDM model. The histograms of (i) $\Delta\chi^{2}(\Lambda{\rm
CDM}-{\rm degen})$ and (ii) $\Delta\chi^{2}({\rm degen}-\Lambda{\rm
CDM})$, corresponding to the difference in goodness-of-fit for the
noise realisations when ${\vec{\bar\xi}}_{\Lambda{\rm CDM}}$ and
${\vec{\bar\xi}}_{\rm degen}$ are used in the template set, are shown
in Fig.\,\ref{chisq}. When the fiducial $\Lambda$CDM model (degenerate
model) is the best-fit and is contained in the template set, values of
$\Delta\chi^{2}$ in Fig.\,\ref{chisq} should be negative. This gives a
measure of our ability to differentiate between models using
correlation functions between redshift slices.  In the first
histogram, when the model for ${\vec{\bar\xi}}_{\Lambda\rm CDM}$ is
contained in the template set, in 95.6\% of cases the noisy
$\Lambda$CDM correlation functions are better fit. Also, when
${\vec{\bar\xi}}_{\rm degen}$ is in the template set, 96\% of the
noisy degenerate correlation functions have better fits. Hence, within
the assumptions we made and in the presence of an intrinsic alignment
signal, these two cosmological models could be
distinguished at the $\sim 2\sigma$-level.

The foregoing example for distinguishing between two cosmological
models should serve as a general illustration only. Owing to our
simplified ansatz for the covariance matrix {\cal C}, which contains
intrinsic ellipticity dispersion only, a more detailed investigation
is not warranted here. In practice, cosmic variance would need to be
taken into account in the covariance matrix when realistic constraints
on cosmological parameters are to be derived. We are currently
investigating ways to obtain a far more realistic representation of
the covariance matrix, to be used in a study of the accuracy of
cosmological parameter determination.

\section{Discussion and conclusions}

It has been suggested that the lensing correlation
function may be contaminated by intrinsic galaxy alignments. 
Since cosmic shear probes the matter power spectrum and enables
constraints to be placed on cosmological parameters such as
$\sigma_{8}$ and $\Omega_{\rm m}$ (e.g. van Waerbeke et al. 2002a), 
it is vital to have the ability to isolate the contribution from 
intrinsic galaxy alignments in order to remove this systematic. Of course, 
intrinsic alignment is interesting in its own
right: its amplitude as a function of physical separation and
its evolution with redshift provides clues about the galaxy formation 
process. 

We have demonstrated that measuring galaxy ellipticity correlation
functions between redshift slices would enable the intrinsic and lensing
contributions to be disentangled. The total signal is decomposed into
template functions, and the fact that intrinsic alignments operate
over a limited physical separation enables the intrinsic component to
be isolated and subracted from the total signal. Our knowledge of the
amplitude of intrinsic alignments is limited, but no strong assumption 
about the behaviour of the intrinsic alignment signal needs to be
made. Here we considered a
modest number of template functions, which can easily be augmented to
cover a wider range of functional forms. For example, any intrinsic
alignment signal arising at the epoch of galaxy formation may be
suppressed by subsequent dynamical interaction, perhaps most pertinent
to galaxy pairs with extremely small physical separations.
In fact, if the reduced $\chi^{2}$ of the best fit is significantly 
larger than 1, this indicates that additional template functions need
to be included.

\begin{figure}
\epsfig{file=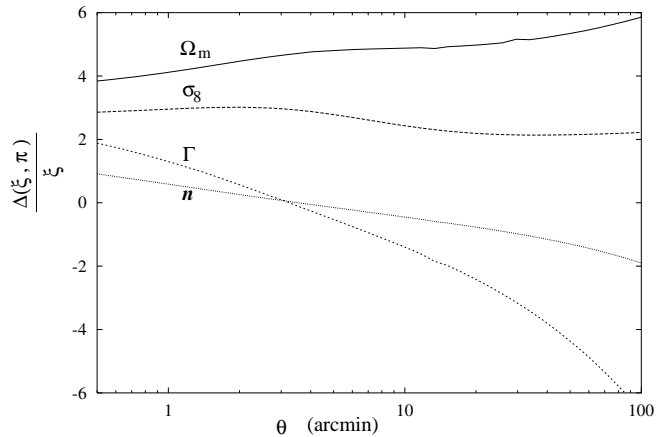, width=88mm}
\caption{
The ratio of partial derivatives of $\xi$ to 
$\xi$ for our fiducial $\Lambda$CDM model, with respect to 
parameters (i) $\Omega_{\rm m}$, (ii) $\sigma_{8}$, (iii) $\Gamma$ and 
(iv) $n$.} 
\label{dxp}
\end{figure}

Our choice of template functions is of course fairly arbitrary. We
have taken functions which have approximately the behaviour 
expected from a cosmic shear measurement. Alternatively, one could
consider a set of generic basis functions, which however, owing to the
dependence on three variables, would require a fairly large set of
functions. Another natural choice of the template functions could be
the following: assuming a reasonable guess for the cosmological model,
characterised by the parameters $\vc \pi_0$, the correlation function
for neighbouring models could be written as
\begin{eqnarray}
&&\bar\xi(\theta,\bar z_i,\bar z_j; \vc \pi)
\approx\\\nonumber
&&\bar\xi(\theta,\bar z_i,\bar z_j; \vc \pi_0)
+(\vc \pi-\vc \pi_0)
{\partial \bar\xi
\over
\partial \vc \pi}(\theta,\bar z_i,\bar z_j,\vc \pi_0)\;,
\end{eqnarray}
and therefore, the set consisting of $\bar\xi(\vc \pi_0)$ and its
partial derivatives with respect to the relevant cosmological
parameters would provide a useful set of template functions. As an
illustration, in Fig.\,\ref{dxp} we have plotted the partial 
derivatives of $\xi(\theta)$ for the case where no redshift 
information is available, i.e. the derivatives of the redshift-averaged 
correlation function, again with ${\ave{\bar z}}\sim 1$. 
For our cosmological model (characterised by $\vc \pi_0$)
we take the fiducial $\Lambda$CDM model. Derivatives are taken with respect 
to (i) $\Omega_{\rm m}$, (ii) $\sigma_{8}$, (iii) $\Gamma$ and (iv) the 
primordial spectral index $n$ (where our fiducial model is 
scale-invariant $n=1$). We plot the ratio
$\frac{\partial\xi}{\partial\pi_{i}}/\xi$, where the numerator
is denoted by $\Delta(\xi,\pi_{i})$.  In the limiting case where
one curve is a scaled version of another,
$\Delta(\xi,\pi_{i})\propto\Delta(\xi,\pi_{j})$, it is
impossible to first order to break the degeneracy between parameters
$\pi_{i}$ and $\pi_{j}$. We again see a nice illustration of the near
degeneracy between $\Omega_{\rm m}$ and $\sigma_{8}$, manifest in the
similarity between $\Delta(\xi,\Omega_{\rm m})$ and
$\Delta(\xi,\sigma_{8})$. Since $\sigma_{8}^{2}$ enters into the
linear power spectrum just as a prefactor, on large angular scales
(i.e. in the linear regime), the curve for $\sigma_{8}$ tends to a
constant, $\Delta(\xi,\sigma_{8})/\xi\to 2 \sigma_8^{-1}$ for
large separations.  Another feature to note is that on the scale of a
few arcminutes the curves for $\Gamma$ and $n$ change sign, implying
that there is less degeneracy between these parameters and either of
$\Omega_{\rm m}$ or $\sigma_{8}$. 

There are several ways in which the method and results discussed
here could be used. One way would be to consider the resulting split into 
intrinsic correlations and lensing signal as the final result, and to 
compare the resulting functions with theories of galaxies formation which 
predict the intrinsic alignment signal, and cosmological models predicting 
the shear correlation function. The resulting fits are, however, difficult
to interpret statistically, i.e. the error bars on the shear
correlation function are difficult to obtain. An alternative would be
to consider the fitted intrinsic signal only, subtract it from the
ellipticity correlation function, and consider the result as the shear
correlation function, together with the corresponding error bars. 
Subsequently, the correlation function can then be used for the
redshift-weighting method of King \& Schneider (2002), of course
yielding much smaller contributions from the intrinsic correlations
than for the unsubtracted data.  Furthermore, the resulting model for
the intrinsic correlation function could also be used as input
for the subtraction method discussed in Heymans \& Heavens (2002).

In addition to providing a key to the suppression of any intrinsic 
alignment signal, photometric redshift estimates enable much tighter
constraints to be placed on cosmological parameters obtained from
cosmic shear surveys, as demonstrated by Hu (1999). Although our
prime goal in this paper is not the constraint of cosmological
parameters, we have illustrated that the degeneracy between 
$\Omega_{\rm m}$ and $\sigma_{8}$ can be lifted by observing
correlation functions between redshift slices, even when an intrinsic 
alignment systematic is present.
   
\begin{acknowledgements}
We would like to thank Marco Lombardi, Patrick Simon, Douglas Clowe, 
and in particular Martin Kilbinger for helpful discussions. 
We would also like to thank the anonymous referee for very helpful
remarks. This work 
was supported by the Deutsche Forschungsgemeinschaft under the project 
SCHN 342/3--1, and by the German Ministry for Science and Education
(BMBF) through the DLR under the project 50 OR 0106.. 
\end{acknowledgements}

\def\ref#1{\bibitem[1998]{}#1}

\end{document}